\documentclass[12pt]{article}
\usepackage[dvips]{graphicx}
\usepackage{latexsym}
\def\slashchar#1{\setbox0=\hbox{$#1$}\dimen0=\wd0%
\setbox1=\hbox{/}\dimen1=\wd1%
\ifdim\dimen0>\dimen1%
\rlap{\hbox to
\dimen0{\hfil/\hfil}}#1\else
\rlap{\hbox to \dimen1{\hfil$#1$\hfil}}/\fi}

\newcommand{\beq}{\begin{equation}}
\newcommand{\eeq}{\end{equation}}
\newcommand{\Frac}[2]{\frac{\displaystyle #1}{\displaystyle #2}}

\newcommand{\gsim}{\stackrel{>}{_\sim}}

\def\mapright#1#2{\smash{
     \mathop{-\!\!\!-\!\!\!-\!\!\!\longrightarrow}\limits^{#1}_{#2}}}

\begin{document}
\thispagestyle{empty}
\begin{titlepage}
\begin{center}
\hfill IFIC/02$-$21 \\ 
\hfill FTUV/02$-$0611 \\
\vspace*{3.5cm} 
\begin{Large}
{\bf QED box amplitude in heavy fermion production} \\[2.25cm]
\end{Large}
{ \sc J. Portol\'es} \ and { \sc P. D. Ruiz-Femen\'\i a}\\[0.5cm]
{\it Departament de F\'\i sica Te\`orica, IFIC, Universitat de Val\`encia -
CSIC\\
 Apt. Correus 22085, E-46071 Val\`encia, Spain }\\[3.5cm]

\begin{abstract}
\noindent
We evaluate the two--photon box contribution to heavy fermion
production in electron positron annihilation, that provides
${\cal O}(\alpha^2)$ electromagnetic corrections to the Born cross
section. The study of its 
non--relativistic expansion, relevant at energies close to the 
threshold of production, is also performed. We also verify that the
threshold expansion of the one--loop integrals correctly reproduces
our results, thus extending the applicability of this technique to 
heavy fermion production diagrams.

\end{abstract}
\end{center}
\vfill
\hspace*{0.5cm} PACS~: 12.20.Ds, 13.40.-f \\
\hspace*{0.5cm} Keywords~: Heavy fermion production, QED, non--relativistic
expansion, asymptotic \\ \hspace*{2.75cm} expansion.
\eject
\end{titlepage}

\pagenumbering{arabic}

\section{Introduction}
\hspace*{0.5cm}
Heavy fermion production processes out of electron positron annihilation,
$e^+ e^-  \rightarrow f \bar{f}$, have become a subject of thorough
study in the last years. Their interest embodies multiple features and 
a wide energy range, from close to threshold production to high--energy
colliders. LEP and LEP2 have provided the appropriate tool pushing behind
this burst. In addition this is among the scattering processes with 
higher expected number of events at a future Linear Collider running
in the $0.5 \, \mbox{TeV} - 1 \, \mbox{TeV}$ energy region like TESLA and
NLC/JLC-X, or CLIC at higher energies. Their interest
arises mainly from the possibility of exploring New Physics and, therefore,
an accurate description within the Standard Model is necessary for the
analyses of data. Projects like ZFITTER \cite{ZFI00} and the ongoing
CalcPHEP \cite{Calc02,Calc03} aim to provide the relevant theoretical
framework for that purpose.
\par
QED corrections seem to be of little interest when probing the quantum
effects within the Standard Model, but it is obvious that their 
contribution, however small, should be considered in order to disentangle
New Physics effects. Besides, if a deeper understanding on the physical
parameters of heavy fermions is intended, electromagnetic $\tau^+ \tau^-$
and heavy quark $Q \overline{Q}$ production out of $e^+ e^-$ annihilation
at threshold energies supplies the required information.
\par
From a theoretical point of view $e^+ e^- \rightarrow f \bar{f}$
cross sections close to threshold evaluated within perturbation theory 
are mislead due to the presence, in the physical system, of a kinematical
variable of the same order than the gauge theory coupling~: the velocity
of the heavy fermion pair in the center of mass of the colliding system,
$\beta = \sqrt{1-4 M^2/s}$, with $M$ the mass of the $f$ fermion. Hence, 
when $\beta \sim \alpha$, care has to be taken in order to resummate terms
as $(\alpha/\beta)^n$ or $(\alpha \ln \beta)^n$ that can give potentially large
contributions \cite{JUAN}. Recently the development of non--relativistic
effective field theories of QED and QCD \cite{CP86} implements the 
suitable systematic procedure to follow. Facilities as the proposed
Tau--Charm Factory, a high--luminosity $e^+ e^-$ collider with a 
center--of--mass energy near the $\tau^+ \tau^-$ production threshold
\cite{TAUC}, would provide excellent information on the mass of this
lepton \cite{taus}. Moreover an accurate determination of the mass of
the top quark (difficult to get at the next hadron colliders) requires
a future lepton collider at the $t \overline{t}$ threshold \cite{TOP}.
Consequently a thorough study of the non--relativistic contribution
to $\sigma (e^+ e^- \rightarrow f \bar{f})$ both from 
electromagnetic and strong interactions is mandatory.
\par
In Ref.~\cite{taus} a detailed study of the threshold behaviour of
$\sigma(e^+ e^- \rightarrow \tau^+ \tau^-)$ was performed, and it was
pointed out that, within the ${\cal O}(\alpha^2)$ electromagnetic
corrections to the Born cross section, the squared amplitude of the
box diagram involving two--photon $\tau^+ \tau^-$ production 
(see Fig.~\ref{fig:box}) had not been considered yet. The electroweak
one--loop contributions to the $e^+ e^- \rightarrow f \bar{f}$
process were evaluated in Ref.~\cite{BMH91}. Here this box contribution
was already taken into account, though an explicit expression was only
given for the $M=0$ case. In this paper we provide the amplitude of
this diagram for a final massive fermion. \footnote{While writing this
article Ref.~\cite{Calc03} appeared. In this preprint a full
expression for the QED box diagram amplitude is also given.}
\par
Once the explicit result is worked out we perform its non--relativistic
expansion in terms of the $\beta$ velocity and we find that the 
contribution of this diagram to the cross section is of 
${\cal O}(\alpha^4 \beta^3)$, that is ${\cal O}(\alpha^2 \beta^2)$
over the Born cross section. The additional suppression driven by
the velocity squared indicates that the contribution of the 
two--photon box diagram to the production of heavy fermions at threshold
is negligible compared to the precision foreseen in the next future.
\par
In Section 2 we detail the calculation of the box diagram contributing
to $e^+ e^- \rightarrow f \bar{f}$ in the limit when 
$m_e \ll M$, and we provide the full analytical result. Section 3 is
dedicated to the study of the threshold behaviour of the box amplitude
as obtained directly from our previous result. We confirm the features of this
threshold amplitude by performing an alternative analysis of the integrals
through the asymptotic expansion method in Section 4.
Our conclusions are collected in Section 5. Finally, 
two Appendices contain the
basic scalar integrals appearing in the article and a comment on the
infrared divergent part of the box amplitude. 

\section{Two--photon box diagram}
\begin{figure}[tb]
\begin{center}
\hspace*{-0.5cm}
\includegraphics[angle=0,width=1.0\textwidth]{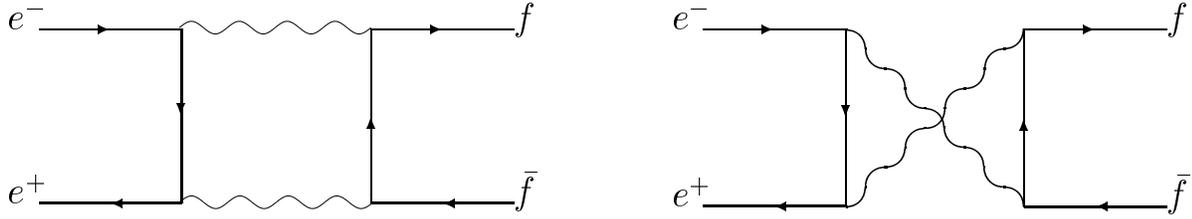}
\caption[]{\label{fig:box} \it Direct (a) and crossed (b) 
box diagrams for $e^+e^-\to f\bar{f}$.}
\end{center}
\end{figure}

\hspace*{0.5cm}
The contribution to the $S$-matrix of the process
$e^-(p) e^+(p^{\prime}) \rightarrow f(k) \bar{f}(k^{\prime})$
of the two--photon box amplitudes is depicted
in Fig.~\ref{fig:box} and it is defined by
\begin{equation}
\label{eq:smatx}
\langle \, f \bar{f} \, | \,  i \, {\cal T} \, | \, e^+ e^- \, \rangle_{box}
=(2\pi)^4\,\delta(p+p^{\prime}-k-k^{\prime})\,i\,{\cal M}_{box} \; 
\; .
\end{equation}

As we are interested in heavy fermion production we will perform
the evaluation for $k^2=k^{\prime 2}=M^2$ and 
$p^2 = p^{\prime 2} = m^2 \, \ll M^2$ (we neglect the electron 
mass where possible). The two--photon box amplitude is gauge 
invariant and, consequently, we perform the calculation by 
taking the Feynman choice. 
The direct box amplitude, Fig.~\ref{fig:box}(a),
is written down following QED Feynman rules as~:
\begin{equation}
{\cal M}_a  =  e^4 Q_f^2 \int\frac{d^4\ell}{i(2\pi)^4}\,
\frac{\{\bar{v}_e(p^{\prime})\,\gamma^{\mu}\,\slashchar{\ell}\,
\gamma^{\nu}u_e(p)\}
\,\{\bar{u}_f(k)\,\gamma_{\nu}\,
(\slashchar{k}-\slashchar{p}+\slashchar{\ell}+M)\,
\gamma_{\mu}\,v_f(k^{\prime})\}}{(\ell^2-m^2)[(\ell+k-p)^2-M^2]
[(\ell-p)^2-\lambda^2]
[(\ell+p^{\prime})^2-\lambda^2 ]}\,\,,
\label{eq:amplitude}
\end{equation}
where we have introduced a photon mass $\lambda$ in order to 
regularize the infrared divergences known to be present in this
amplitude.
The crossed box diagram in Fig.~\ref{fig:box}(b)
can be obtained from (\ref{eq:amplitude}) by adding an overall minus
sign, reversing the order of the 
$\gamma_{\mu}$, $\gamma_{\nu}$ matrices in the heavy fermion bilinear,
and performing the 
substitutions $k \rightarrow k^{\prime}$ everywhere (except for the
spinors) and $M \rightarrow - M$. Hence, in 
Eq.~(\ref{eq:smatx}), ${\cal M}_{box} \, = \, {\cal M}_a + {\cal M}_b$.
The evaluation of the integrals is slightly 
cumbersome but straightforward and the details are given in the Appendix A.
\par
With the definition of the Mandelstam variables $s = (p+p^{\prime})^2$
and $t = (p-k)^2$, the spinorial structure of ${\cal M}_{box}$ is decomposed 
into 4 sets of amplitudes $L_{i}^{\rho \kappa}$ multiplied
by corresponding coefficients $w_i^{\rho}$:
\begin{equation}
{\cal M}_{box} (\kappa;s,t)= e^4Q_f^2\sum_{i=1}^{4}\sum_{\rho = \pm 1} 
L_i^{\rho \kappa} \,w_i^{\rho} \; \; ,
\label{eq:finalM}
\end{equation}
with the basic amplitudes
\begin{eqnarray}
L_1^{\rho \kappa} & = & [ \, \bar{v}_e(p^{\prime})\,
\gamma_{\mu}\,P_{\kappa}\,u_e(p)\,]
\,\,[\, \bar{u}_f(k)\,\gamma^{\mu}\,(1+\kappa\rho\,\gamma_5)\,v_f(k^{\prime})\,]
\; ,
\nonumber\\[3mm]
L_2^{\rho \kappa} & = & [ \, \bar{v}_e(p^{\prime})\,
\slashchar{k}\,P_{\kappa}\,u_e(p) \,]
\,\,[\, \bar{u}_f(k)\,\slashchar{p}\,(1+\kappa\rho\,\gamma_5)\,v_f(k^{\prime}) \,]
\; ,
\nonumber\\[3mm]
L_3^{\rho \kappa} & = & [ \, \bar{v}_e(p^{\prime})\,
\slashchar{k}\,P_{\kappa}\,u_e(p) \,]
\,\,[\, \bar{u}_f(k)\,(1+\kappa\rho\,\gamma_5)\,v_f(k^{\prime}) \,] \; ,
\nonumber\\[3mm]
L_4^{\rho \kappa} & = & [ \, \bar{v}_e(p^{\prime})\,
\gamma_{\mu}\,P_{\kappa}\,u_e(p) \,]
\,\,[\, \bar{u}_f(k)\gamma^{\mu}\,\slashchar{p}\,(1+\kappa\rho\,\gamma_5)\,v_f(k^{\prime}) \,]
\,\,.
\label{eq:spinorbasis}
\end{eqnarray}
The latter have been written in terms of the initial state $e^+e^-$
chiral projectors
\begin{equation}
P_{\kappa}=\frac{1}{2}\Big( 1+\kappa \,\gamma_5 \Big)\,\,\,,\,\,\,\kappa =\pm 1
\,\,,
\end{equation}
which, as we are considering massless initial fermions, satisfy 
$P_{\kappa} u_e(p) = u_e(p)$, being $\kappa$ the initial
electron helicity; in the massless limit, positron helicity is forced to 
be $-\kappa$ to have a non vanishing amplitude. The dependence of 
${\cal M}_{box}$ on the spin state of the final state fermions 
has not been explicitly stated. 
\par
The $w_i^{\rho}$ coefficients can be written in terms
of four auxiliary functions ${\cal F}_{i}, \; \, i=0,1,2,3$~:
\begin{eqnarray}
w_1^+ & = & \Frac{1}{2}{\cal F}_0(s,t)\,\,, \nonumber \\[2mm]
w_1^- & = & - \, \Frac{1}{2} {\cal F}_0(s,u)\,\,, \nonumber \\[2mm]
w_2^+ & = & {\cal F}_1(s,t)+{\cal F}_2(s,t)\,\,, \nonumber \\[2mm]
w_2^- & = & {\cal F}_1(s,u)+{\cal F}_2(s,u)\,\,, \nonumber \\[2mm]
w_3^+ & = & M\Big({\cal F}_1(s,u)-{\cal F}_1(s,t)+
{\cal F}_3(s,u)-{\cal F}_3(s,t) \Big)\,\,,
\label{eq:www}\\[2mm]
w_3^- & = & M\Big({\cal F}_3(s,u)-{\cal F}_3(s,t) \Big) \,\,,
\nonumber \\[2mm]
w_4^+ & = & - \, \Frac{1}{2} M\Big({\cal F}_2(s,t)-{\cal F}_2(s,u)\Big)\,\,,
\nonumber \\[2mm]
w_4^- & = &  \, \Frac{1}{2} M\Big({\cal F}_2(s,t)-{\cal F}_2(s,u)\Big)\,\,,
\nonumber 
\end{eqnarray}
that read
\begin{eqnarray} 
{\cal F}_0(s,t) &=& \,{\cal F}_0^{\lambda}(s,t) \, + \, 
\frac{M^2-t}{(M^2-t)^2+st}
\bigg\{ \, t \, \left( \, s  D_0 \, - \, 2  C_t \, \right)
 -  M^2 \, \frac{M^2-s-t}{M^2-t} \left( \, s  \overline{D}_0 \, - \, 2 
\overline{C}_t \, \right) \label{eq:fifunc1}\\[3mm] & & 
\; \; \; \; \; \; \; \; \; \; \; \; \; \; \; \; \; \; \; \; \; \; \; 
\; \; \; \; \; \; \; \; \; \; \; \; \; \; \;\; \; \; \; \;
- \, \left( 2M^2-s-2t  \right)  \left( C_s + C_M \right) \bigg\}
\,\,,\nonumber\\[6mm]
{\cal F}_1(s,t) &=&
\,\frac{1}{(M^2-t)^2+st} \, 
\bigg\{ \, 2 (B_t -  B_s) + \Frac{4 M^2}{M^2-t} \,  (B_M - B_t) \,
\label{eq:fifunc2}\\[3mm]
&& \; \; \; \; \; \; \; \; \; \; \; \; \; \; \; \; \; \; \; \; \; \; \; 
+ \, \frac{M^2-t}{(M^2-t)^2+st} \bigg[ \, (2M^2-s-2t) \, 
 \left((M^2-t)(s D_0 - 2 C_t) 
+ s \, C_s \, \right)  \nonumber \\[3mm]
&& \; \; \; \; \; \; \; \; \; \; \; \; \; \; \; \; \; \; \; \; \; \; \; \; \; 
\; \; \; \; \; \; \; \; \; \; \; \; \; \; \; \; \; \; \; \; \; \; \; \; \;  
\;  + \, \left( \,  2 (M^2-t)^2+s(2t+s-4M^2) \, \right) 
C_M \, \bigg] \, \bigg\} \; \; , \nonumber \\[6mm]
{\cal F}_2(s,t) &=& \, {\cal F}_2^{\lambda}(s,t) \, - \, 
\,\frac{1}{(M^2-t)^2+st}\, \bigg\{ \, (M^2+t) \, \left( \, s D_0 \, - \, 
2 C_t \, \right) \, - \, (s-4M^2) \, C_M
\label{eq:fifunc3}\\[3mm]
&& \; \; \; \; \; \; \; \; \; \; \; \; \; \; \; \; \; \; \; \; \; \; \; \; \;
\; \; \; \; \; \; \; \; \; \;\; \; \; \; \; \; \; \; \; \; \; 
- \, (2M^2-s-2t) \, C_s \bigg\}\,\,,\nonumber\\[6mm]
{\cal F}_3(s,t) &=&
\,\frac{1}{(M^2-t)^2+st} \, 
\bigg\{ \, \Frac{2t}{M^2-t} \,  (B_t -  B_M) \, + \, 
 \Frac{2(M^2+t)}{4M^2-s} \, (B_s - B_M) \,
\label{eq:fifunc4}\\[3mm]
&& \; \; \; \; \; \; \; \; \; \; \; \; \; \; \; \; \; \; \; \; \; \; \;
- \, \frac{(M^2-t)^2}{(M^2-t)^2+st} \bigg[ \, (M^2-t) \, ( s D_0 - 2 C_t) 
\, + \,  s \, C_s \, 
\bigg]
\nonumber \\[3mm]
&& \; \; \; \; \; \; \; \; \; \; \; \; \; \; \; \; \; \; \; \; \; \; \; 
+ \, 
\Frac{1}{(4M^2-s)((M^2-t)^2+st)}
\left[-4\, M^8 \, + \, 6(s+2t)\, M^6 \right. \nonumber \\
&& \; \; \; \; \; \; \; \; \; \; \; \; \; \; \; \; \; \; \; \; \; \; \; \; \; 
\; \; \; \; \; \; \; \; \; \; \; \; \; \; \; \; \; \; \; \; \; \; \; 
\; \; \; \; \; \; \; \; \; \;\; \; \; \; \; \; \; \; \; \;\; 
\left. - \, (s+2t)\, (s+6t) \, M^4 \right. \nonumber \\
&& \; \; \; \; \; \; \; \; \; \; \; \; \; \; \; \; \; \; \; \; \; \; \; \; \; 
\; \; \; \; \; \; \; \; \; \; \; \; \; \; \; \; \; \; \; \; \; \; \; 
\; \; \; \; \; \; \; \; \; \;\; \; \; \; \; \; \; \; \; \;\; 
\left. + \,2t  \, (s^2+ts+2t^2) \, M^2 \, +
\, s^2t^2 \, \right]
 C_M \, \bigg\} \; \; . \nonumber 
\end{eqnarray}
Full expressions for the scalar functions $B_s$, $B_t$, $B_M$,  $C_s$, 
$C_t$, $\overline{C}_t$, $C_M$, $D_0$ and $\overline{D}_0$ can be
found in Eqs.~(\ref{eq:D0}-\ref{eq:BM}) of the Appendix A. It
can be seen from ${\cal F}_1(s,t)$ and ${\cal F}_3(s,t)$ that the  
two--point functions $B_t$, $B_s$ and $B_M$ only appear in non--divergent
combinations while the rest of scalar integrals in ${\cal F}_i(s,t)$
are UV finite. Clearly ${\cal M}_{box}(\kappa; s,t)$
is ultraviolet finite.
Scalar integrals $D_0$ ($\overline{D}_0$) and $C_t$
($\overline{C}_t$) are
infrared divergent for vanishing photon mass $\lambda$, however the 
combinations $s D_0 - 2 C_t$  and $s \overline{D}_0 - 2 \overline{C}_t$ 
are divergenceless. Hence all the divergences
in the ${\cal F}_i(s,t)$ functions are collected in ${\cal F}_0^{\lambda}(s,t)$
and ${\cal F}_2^{\lambda}(s,t)$, given by~:
\begin{eqnarray} \label{eq:f02IR}
{\cal F}_0^{\lambda}(s,t) \, & = & \, 2 \, \left(  M^2  -  s  -  t
 \right) \, \overline{D}_0 \; \; , \nonumber \\[3mm]
{\cal F}_2^{\lambda}(s,t) \, & = & \, 2 \, D_0 \; \; ,
\end{eqnarray}
and we get for the infrared divergent part of the box amplitude~:
\begin{equation} \label{eq:boxIR}
{\cal M}_{box}^{IR} \, = \, \Frac{e^4 Q_f^2}{8 \pi^2 \, s} \, \left(
L_1^{+ \kappa} \, + \, L_1^{- \kappa} \right) \, 
\ln \left( \Frac{M^2-u}{M^2-t} \right) \, \ln \left(
\Frac{-s-i \delta}{\lambda^2} \right) \; \; \; . 
\end{equation}
A more complete discussion on the infrared structure of the QED box diagram
and the determination of ${\cal M}_{box}^{IR}$ is relegated to Appendix B.
\par
Incidentally our result can be used to evaluate the similar
two--gluon box contribution to the heavy quark production out
of light quarks, $q(A) \bar{q}(B) \rightarrow Q(C) \overline{Q}(D)$
(between parentheses we label the colour quantum numbers). In order
to get this amplitude we need to substitute the $e^4 Q_f^2$ factor 
in Eq.~(\ref{eq:finalM}) by 
$g_s^4 (t^b t^a)_{BA} \{t^a,t^b\}_{CD}$ (a sum over repeated indices is
implied). \footnote{In this colour factor $t^i = \lambda^i/2$, where
$\lambda^i$ are the $SU(3)$ Gell--Mann matrices and 
$Tr(t^i t^j) = \delta^{ij}/2$.}
\par
We have checked that our amplitude in Eq.~(\ref{eq:finalM}), 
when summed over polarizations,
coincides with a recent result found in 
Ref.~\cite{Calc03}, 
though these authors use a different basis of spinor operators.  
Moreover, from our calculation for ${\cal M}_{box}$, we can recover 
the case where the final fermions are massless. 
The limit $M\to 0$ can be
directly applied to the $w_i^{\pm}$ coefficients, Eqs.~(\ref{eq:www}),
and to the scalar integrals quoted in the Appendix A. Within this limit
our result agrees with the earlier calculation in Ref.~\cite{BMH91}.

\section{Heavy fermion production at threshold}
\hspace*{0.5cm}
Close to $f\bar{f}$ threshold, it is more convenient to 
expand the production 
amplitude in terms of the fermions velocity in the 
center--of--mass of the colliding system $\beta=\sqrt{1-4M^2/s}$. 
Hence production amplitudes are written in a combined
expansion in powers of $\alpha$ and $\beta$, and the importance
of each contribution is estimated taking $\alpha\sim \beta$. This
feature spoils the perturbative expansion in QED due to the
appearance of ${\cal O}(\alpha^n/\beta^n)$ and 
${\cal O} (\alpha^m ln^n \beta)$ terms that diverge as $\beta \rightarrow 0$.
As a consequence, a resummation of such terms is necessary to avoid
a breakdown of the perturbative series, and  
well-known results from the familiar non-relativistic 
quantum mechanics are obtained. Nevertheless
it is somewhat misleading to associate the
appearance of these Coulomb terms to the non-relativistic 
motion of the fermion pair, as the scattering amplitude calculated 
from quantum mechanics does not show any kinematic singularity close
to threshold: their ultimate origin is the inadequacy of
the diagrammatic QED expansion in powers of $\alpha$ to account for
the correct non-relativistic dynamics. Keeping this in mind, one should
not discard, a priori, divergent terms in the velocity appearing 
in any QED diagram involving fermions with small velocities.
\par
In Ref.~\cite{taus} it was pointed out that the contribution at threshold
of the two--photon box diagram should be analysed in a NNLO calculation
of $\sigma(e^+e^-\to \tau^+\tau^-)$. 
In this Section we proceed to perform the expansion on 
${\cal M}_{box}$ as given in Eq.~(\ref{eq:finalM}).
The leading terms in the velocity expansion
of the coefficients $w_i^{\pm}$ can be obtained by taking into account
the dependence of the Mandelstam
invariants $s,t,u$ on the velocity $\beta$ and the angle $\theta$
between the momenta of the heavy fermion and the electron in the 
colliding center--of--mass system. The relation is given by~: 
\begin{equation}
s= \frac{4M^2}{1-\beta^2}\,\,\,\; \; , \;  \; \,\,\,t=M^2-\frac{2M^2}{1-\beta^2}
(1-\beta\cos \theta)\,\,\; \;  \,,\,\; \; \,\,u=M^2-\frac{2M^2}{1-\beta^2}
(1+\beta\cos \theta)\,\,.
\label{eq:inv}
\end{equation} 
Carrying these expressions to the $w_i^{\pm}$ coefficients displayed in
Eq.~(\ref{eq:www}) and neglecting ${\cal O}(\beta^2)$ terms
 we obtain~:
\begin{eqnarray}
w_1^+ & = &\frac{1}{384M^2\pi^2} \bigg[  
-\pi^2 + 3 \ln^2 
\frac{4 M^2}{\lambda^2} -3 \ln^2 \frac{m^2}{\lambda^2}
\nonumber\\[3mm]
&& \; \; \; \; \; \; \; \; \; \; \; \; \; \; \; \; 
 +\, \bigg(8-14 i \pi  -8 \ln 2 +
12 \ln \frac{4 M^2}{\lambda^2} \bigg)\,\beta\,\cos\theta  \,\bigg]
\, + \, {\cal O}(\beta^2) \; \; ,
\nonumber \\[4mm]
w_1^- & = & -w_1^+(\beta \to -\beta)\,\, \; ,
\nonumber\\[4mm]
w_2^+ & = &\ \frac{1}{384M^4\pi^2} \bigg[\,  
\pi^2-8+8 i \pi  + 8 \ln 2  - 3 \ln^2 
\frac{4 M^2}{\lambda^2} + 3 \ln^2 \frac{m^2}{\lambda^2} \,\, 
\nonumber\\[3mm]
&& \; \; \; \; \; \; \; \; \; \; \; \; \; \; \; \;  
+ \, \bigg( \pi^2 - 34 + 4 i \pi + 16 \ln 2 + 
12 \ln  \Frac{4 M^2}{\lambda^2} 
- 3 \ln^2 
\frac{4 M^2}{\lambda^2} + 3 \ln^2 \frac{m^2}{\lambda^2} \bigg) \, 
\beta \cos\theta \, \bigg]
\nonumber\\[3mm]
&& \; \; \; \; \; \; \; \; \; \; \; \; \; \; \; \; 
+ \;  {\cal O}(\beta^2)\,\,,
\nonumber \\[4mm]
w_2^- & = & w_2^+(\beta \to -\beta)\,\,,
\label{eq:wbeta}\\[4mm]
w_3^+ & = & \frac{1}{240M^3\pi^2} \,\bigg(  
37+2 i \pi  -64 \ln 2  \,\bigg)\,\beta\cos\theta \, + \, {\cal O}(\beta^2)\,\,,
\nonumber\\[4mm]
w_3^- & = & \frac{-1}{480M^3\pi^2} \,\bigg( 
11+ i\pi  -32 \ln 2  \,\bigg)\,\beta\cos\theta \, + \, {\cal O}(\beta^2) \,\,,
\nonumber\\[4mm]
w_4^+ & = & \frac{-1}{384M^3\pi^2} \,\bigg(  
\pi^2+6i\pi  -48 \ln 2 + 
12 \ln  \Frac{4 M^2}{\lambda^2} 
- 3 \ln^2 
\frac{4 M^2}{\lambda^2} + 3 \ln^2 \frac{m^2}{\lambda^2} \bigg) \, \beta 
\cos\theta \, + \, {\cal O}(\beta^2) \, ,
\nonumber\\[4mm]
w_4^- & = &-w_4^+\,\,.
\nonumber
\end{eqnarray}
The amplitudes $L_i^{\rho \kappa}$, containing fermion wave functions,
must also be
expanded in terms of $\beta$ to fulfill the expansion of ${\cal M}_{box}$
at small velocities. We shall not give the full result of such expansion, but
just quote their leading behaviour, which can be easily obtained by choosing an
explicit representation of the gamma matrices and spinors. We thus get:
\begin{equation}
L_1^{\rho \kappa}={\cal O}(1)\, ,\,\; \; \; \, \; \; \;
 \,\,\,L_2^{\rho \kappa}={\cal O}(\beta)\, ,
\; \; \; \,\,\,\,\; \; \;  \,
L_3^{\rho \kappa}={\cal O}(\beta)\,,\,\; \; \; \,\,\; \; \; 
\,\, \,L_4^{\rho \kappa}={\cal O}(1)
\,\,.
\label{eq:Ls}
\end{equation} 
The terms quoted in Eqs.~(\ref{eq:wbeta}) together
with the expansion in Eq.~(\ref{eq:Ls}) allow us to obtain the leading  
near threshold contribution to the cross section 
of the box amplitude ${\cal M}_{box}$.
Recall that, by virtue of Furry's theorem, the interference of the QED
box amplitude
with other one-loop amplitudes for the process $e^+e^-\to f\bar{f}$ 
vanishes and, consequently, $|{\cal M}_{box}|^2$
adds incoherently to the rest of ${\cal O}(\alpha^4)$
corrections to $\sigma(e^+e^-\to f\bar{f})$,
as studied in Ref.~\cite{taus}.
The final result for the squared and averaged box amplitude is~:
\begin{eqnarray}
\frac{1}{4}\sum_{pol.}|{\cal M}_{box}|^2 =  \left( Q_f \,\alpha \right)^4
\!\!\!\!&\Bigg\{&\!\!\!\!  \frac{16}{9}\Big( \pi^2+(1-\ln 2)^2 \Big)
\label{eq:M2}\\[3mm]
\!\!\!\!  \!+ &&\!\!\!\! \! \! \! \, \! \! \! \! \! \! \!
\Bigg[-\frac{1}{2}L_M^4-4L_M^3 -2L_M^3 \ell_m +
\bigg(-2\ell_m^2-12\ell_m+
\frac{8}{3} \ln 2 +\frac{\pi^2}{3}+\frac{160}{3}\bigg)\,L_M^2
\nonumber\\[3mm]\!\!\!\!&&\!\!\!\!  \!\!\! +
\,\bigg(\!\! -8\ell_m^2+\Big(\frac{16}{3} \ln 2+\frac{2}{3} \pi^2
+\frac{320}{3}\Big)\ell_m
-288\ln 2+\frac{4}{3} \pi^2+32\bigg)\,L_M
\nonumber\\[3mm]\!\!\!\!&&\!\!\!\! \!\!\!+ \,
56 \, \ell^2_m+\Big(-288\ln 2 +\frac{4}{3} \pi^2+32\Big)\ell_m
+\frac{3088}{9}\ln^2 2 -\frac{800}{9}\ln 2
\nonumber\\[3mm]\!\!\!\!&&\!\!\!\!\!\!\!
-\frac{\pi^4}{18}
-\frac{8}{9}\pi^2 \ln 2-\frac{14}{3} \pi^2+\frac{16}{9}\,\Bigg]
\cos^2\theta \, \Bigg\}\,\beta^2
\, + \, {\cal O}(\beta^3)\,\,,
\nonumber
\end{eqnarray}
with
\begin{equation}
L_M\equiv\ln \frac{4M^2}{m^2}\qquad\qquad
\mbox{and}\qquad\qquad
\ell_m\equiv\ln \frac{m^2}{\lambda^2}
\qquad.
\label{eq:Lm}
\end{equation} 
Hence we conclude that 
the result in Eq.~(\ref{eq:M2}), proportional to $ \alpha^4\beta^2$,
represents a N$^4$LO correction with respect the LO result 
(the tree level $e^+e^-\to f \bar{f}$ amplitude squared, which
is already of ${\cal O}(\alpha^2)$). In Ref.~\cite{taus},
box amplitudes were not included with the rest of the one-loop 
diagrams to complete the NNLO calculation of
$\sigma(e^+e^-\to \tau^+\tau^-)$ at threshold, their behaviour with
$\beta$ being unknown. Our evaluation 
of $|{\cal M}_{box}|^2$ has proven that this is, indeed, $\beta^2$
suppressed with respect the NNLO contributions considered in 
\cite{taus}. 

\section{Threshold amplitude by asymptotic expansion of integrals}
\hspace*{0.5cm}
The counting of powers of the velocity appearing in a defined
amplitude is not straight because $\beta$ is not a parameter in the
Lagrangian, but rather a dynamic scale which is driven by the 
propagators inside loop integrals. In recent years, this 
issue made awkward to define a non-relativistic effective theory 
suitable for describing quarks and leptons at low velocities. 
Important progress was made after the development of the
threshold expansion by Beneke and Smirnov \cite{threshold}. This 
technique allows for an asymptotic expansion of Feynman integrals
near threshold, providing a set of much simpler integrals which
are manifestly homogeneous in the expansion parameter and so 
have a definite power counting in the velocity. The procedure
should confirm that the two--photon box amplitude is not 
enhanced at low $\beta$, as we have found by explicit evaluation.
This we discuss in the following.
\par
The expansion method, described in Ref.~\cite{threshold}, begins by identifying 
the relevant momentum regions in the loop integrals, which follow from the
singularity structure of the Feynman propagators dictated by the 
relevant scales that appear in the problem. For on-shell
scattering amplitudes of heavy fermions, three scales are identified: the
heavy fermions mass, $M$, their relative 3-momentum, 
$|{\mbox{\boldmath $k$}}|\sim M\beta$ and their energy $k_0\sim M\beta^2$.
Accordingly, the loop
four momentum near the singularities can be in any of the following
regimes:
\begin{eqnarray}
hard\ :&&  \ell_0 \, \sim \, |{\mbox{\boldmath $\ell$}}| \, \sim \, M\,,
\nonumber\\[3mm]
soft\ :&&  \ell_0 \, \sim  \, |{\mbox{\boldmath $\ell$}}| \, \sim \, M\beta\,,
\nonumber\\[3mm]
potential\ :&& \ell_0 \, \sim \, M\beta^2\; \, , \, \;
|{\mbox{\boldmath $\ell$}}| \, \sim M\beta\,,
\nonumber\\[3mm]
ultrasoft\ :&& \ell_0 \, \sim \, 
|{\mbox{\boldmath $\ell$}}| \, \sim \,  M\beta^2\,.
\label{eq:regimes}
\end{eqnarray}    
The original integral is then decomposed into a set
of integrals, one for every region, and a Taylor 
expansion in the parameters, which are small in each regime, is performed. 
Every integral, containing just one scale, will thus contribute 
only to a single power in the velocity 
expansion. The procedure requires the use of dimensional regularization
in handling the integrals,
even if they are finite, in order to assure that the result from
each regime just picks up the corresponding pole contribution and 
vanishes outside. Following this heuristic rules, the authors of 
Ref.~\cite{threshold} reproduce the exact $\beta$ expansion of some one-loop
and two-loop examples. Although a formal proof of the validity of
the asymptotic expansion close to threshold has not been given, the perfect
agreement in the examples supports their use in general one-loop diagrams.
We provide a new test by addressing the rules to the QED box amplitude with
$e^+e^-$ in the initial state, extending the use of the threshold
expansion to diagrams with heavy and massless fermions in the external legs
(i.e. production-like diagrams). We will keep the electron mass finite along
the procedure, although much smaller than any other scale, to keep track 
of the logarithms of $m$ present in the box amplitude.
\par
Our amplitude ${\cal M}_{box}$ is characterized, as shown in the
Appendix A, by the four point
integrals $D_0,D_{\mu},D_{\mu\nu}$ in (\ref{eq:Ds}). If 
present, inverse powers of the velocity in ${\cal M}_{box}$ can only originate
from these integrals. In addition, we can focus on the behaviour of the scalar 
integral $D_0$, as the $\ell_{\mu},\ell_{\mu}\ell_{\nu}$ vectors in 
$D_{\mu}$ and $D_{\mu\nu}$ will produce factors
of one of the scales of the problem ($M,M\beta$ or $M\beta^2$) in the 
numerator of the amplitude without affecting the leading singular behaviour
in $\beta$.
Let us change the routing of momenta 
in $D_0$ (\ref{eq:Ds}) in order to make the scaling arguments more
transparent:
\begin{equation}
\! \! \! \! \! \,  D_0  =   \int  \Frac{d^D \ell}{i (2 \pi)^D} \, 
\frac{1}{[(Q/2+T/2-\ell)^2-m^2][(Q/2+R/2-\ell)^2-M^2]
[\ell^2-\lambda^2]
[(Q-\ell)^2-\lambda^2 ]}\,,
\label{eq:D0routed}
\end{equation}
where the standard $+i\delta$ prescriptions are implicitly understood in the
propagators, the $Q$ and $R$ vectors are defined in 
relation with Eq.~(\ref{eq:Ds}) and $T = p - p^{\prime}$. 
The external four vectors $Q$ and $R$ scale as $M$ and $M\beta$
respectively, while 
$T^2=-s+4m^2 \sim M^2$. 
Using momentum $T$ is preferred to the electron (positron) momentum $p$ 
($p^{\prime}$) because, the spatial and time components
of the latter, although scale as $M$, cancel out in the total momentum squared
$p^2=m^2\sim 0$.
The infrared regularization of the integrals is automatically guaranteed 
by dimensional regularization and, therefore, we will not longer retain
a fictitious mass for the photon.
\par
In the potential region $\ell_0 \ll |{\mbox{\boldmath $\ell$}}| \ll M$ and,
accordingly, we can expand terms in the propagators. 
The leading contribution is
\begin{equation}
D_0^p  =  \int \Frac{d^D \ell}{i (2 \pi)^D} \,
\frac{1}{({\mbox{\boldmath $\ell$}}\cdot{\mbox{\boldmath $T$}})\,
(-{\mbox{\boldmath $\ell$}^2}+{\mbox{\boldmath $\ell$}}
\cdot{\mbox{\boldmath $R$}}-Q_0\ell_0)\,
(-{\mbox{\boldmath $\ell$}^2})\,
(Q_0^2)}\,\,,
\label{eq:potential}
\end{equation} 
where we have also dropped the term $-{\mbox{\boldmath $\ell$}^2}$ in the
electron propagator to be compared to 
${\mbox{\boldmath $\ell$}}\cdot{\mbox{\boldmath $T$}}\sim
M^2\beta$. 
The overall scaling of the potential integration is easily estimated to be of
order $M^4\beta^5/M^8\beta^5\sim 1/M^4$, so no velocity enhancement is this
region is expected. In fact, the integral above is zero because,
closing the $\ell_0$ integration contour in the lower half-plane, the pole at
$\ell_0=({\mbox{\boldmath $\ell$}}\cdot{\mbox{\boldmath $R$}}
-{\mbox{\boldmath $\ell$}}^2)/Q_0+i\delta$ lies outside
\footnote{Notice that the $\ell_0$--integration in $D_0^p$ does not
vanish in the outer semicircle. Rigorously we should keep the $\ell_0^2$
term in the heavy fermion propagator, so $D_0^p$ is well defined. Poles would
then be located at $\ell_0^{\pm} = \frac{1}{2}\Big( Q_0 \pm 
\sqrt{Q_0^2-4({\mbox{\boldmath $\ell$}}
\cdot{\mbox{\boldmath $R$}}-{\mbox{\boldmath $\ell$}}^2)-i \delta
}\Big)$. The root $\ell_0^+$ scales as $M$ and is taken into account 
in the hard region
while $\ell_0^-
=({\mbox{\boldmath $\ell$}}\cdot{\mbox{\boldmath $R$}}
-{\mbox{\boldmath $\ell$}}^2)/Q_0+i\delta$ once we consider that
$|{\mbox{\boldmath $\ell$}}| \ll M$ in the potential region, and we recover the
above result.}. Similarly, subleading
terms in the expansion of propagators in this region are vanishing, as they
share the same pole structure.
\par
When the loop momentum $\ell$ is soft or ultrasoft, the assumption
$\ell_0\sim |{\mbox{\boldmath $\ell$}}| \ll M$ leads to the same expansion
of the propagators in $D_0$:
\begin{equation}
D_0^{s,us}  =  \int \Frac{d^D \ell}{i (2 \pi)^D} \,
\frac{1}{({\mbox{\boldmath $\ell$}}\cdot{\mbox{\boldmath $T$}}-Q_0\ell_0)\,
(-Q_0\ell_0)\,
(\ell_0^2-{\mbox{\boldmath $\ell$}^2})\,
(Q_0^2)}\,\, . 
\label{eq:soft-us}
\end{equation} 
It scales as $1/M^4$ in both the soft and ultrasoft regimes and, indeed,
vanishes in dimensional regularization because, after picking up the residue
in the lower plane, $\ell_0=|{\mbox{\boldmath $\ell$}}|-i\delta$,
the remaining $D-1$ dimension integral is scaleless:
\begin{eqnarray}
D_0^{s,us}  &=& \frac{1}{2Q_0^3}
\int \Frac{d^{D-1} \ell}{(2 \pi)^{D-1}} \,
\frac{1}{|{\mbox{\boldmath $\ell$}}|^2}
\frac{1}{(Q_0|{\mbox{\boldmath $\ell$}}|-
{\mbox{\boldmath $T$}}\cdot{\mbox{\boldmath $\ell$}})} 
\nonumber\\[3mm]
&=&\frac{1}{2Q_0^3}
\int \frac{d\Omega_{D-1}}{(2 \pi)^{D-1}}\frac{1}
{(Q_0-|{\mbox{\boldmath $T$}}|\cos {\varphi})}
\int d|{\mbox{\boldmath $\ell$}}| |{\mbox{\boldmath $\ell$}}|^{D-2}
\frac{1}{|{\mbox{\boldmath $\ell$}}|^3}\,=\,0 \,\,,
\label{eq:soft-usvanish}
\end{eqnarray}
with ${\varphi}$ the angle between the vectors ${\mbox{\boldmath $T$}}$
and ${\mbox{\boldmath $\ell$}}$. The same argument holds for subleading
terms in this region. 
\par
Finally, the integral in
the hard region is obtained by dropping out terms involving
non-relativistic fermion three-momenta from propagators. Hence, the
only scale which
remains is the hard parameter $M$, and so there is no additional
velocity dependence in the denominators. More explicitly, 
the expanded integral in the
hard regime, at leading order in $\beta$, is
\begin{equation}
D_0^{h,{\cal O}(1)}  =  \,\int \Frac{d^D \ell}{i (2 \pi)^D} \,
\frac{1}{(\ell^2- \ell \cdot T -Q\cdot\ell)\,
(\ell^2-Q\cdot\ell)\,
\ell^2 \,
(Q-\ell)^2}\,\,,
\label{eq:hard}
\end{equation}
and there is no need to separate time from spatial components in the
integration. The above integral trivially scales a $1/M^4$, and its explicit 
calculation in $D=4-2\epsilon$ dimensions
has been performed following Ref.~\cite{kosower}:
\begin{equation}
D_0^{h,{\cal O}(1)} \, = \, \frac{\mu^{-2\epsilon}}{8\pi s^2}\,
\ln\frac{s}{m^2} \,\left[ \frac{1}{\epsilon}
-\ln \left(\Frac{-s-i\delta}{\mu^2} \right) +
\ln (4\pi) - \gamma_E \right]\,\,,
\label{eq:hardeval}
\end{equation} 
where
$\gamma_E$ is the Euler-Mascheroni constant. Terms proportional to the
electron mass $m$ have been dropped. The pole in Eq.~(\ref{eq:hardeval})
is of infrared origin, and
it is the analogous to the $\ln \lambda^2$ term in the full result of $D_0$,
Eq.~(\ref{eq:D0}). Indeed, Eq.~(\ref{eq:hardeval}) reproduces the leading term
in the velocity expansion of $D_0$, after the usual replacement
$\ln \lambda^2 \to (4\pi)^{\epsilon}/\Gamma(1-\epsilon)/\epsilon$.
\par
The following order in the expansion within the 
hard region would have a $\ell \cdot R=-{\mbox{\boldmath $\ell$}} 
\cdot{\mbox{\boldmath $R$}}$ term in the numerator, and it would behave
as $\beta/M^4$:
\begin{eqnarray}
D_0^{h,\,{\cal O}(\beta)} & = &  \, \int \Frac{d^D \ell}{i (2 \pi)^D} \,
\frac{\ell\cdot R}{(\ell^2- \ell \cdot T -Q\cdot\ell)\,
(\ell^2-Q\cdot\ell)^2\,
\ell^2 \,
(Q-\ell)^2}
\nonumber\\[3mm]
& = & \frac{R\cdot T}{T^2}\,\left( D_0^h- 
\int \Frac{d^D \ell}{i (2 \pi)^D} \,
\frac{1}{(\ell^2-Q\cdot\ell)^2\,\ell^2 \,(Q-\ell)^2}\right)
\nonumber\\[3mm]
& = &  \frac{\beta\cos\theta}{8\pi^2 s^2}\,\mu^{-2 \epsilon}
\left(\ln\frac{s}{m^2}-2 \right)
\left[\frac{1}{\epsilon}
-\ln \left(\Frac{-s-i\delta}{\mu^2} \right) +
\ln (4\pi) - \gamma_E \right]\,\,,
\label{eq:hard2eval}
\end{eqnarray}
which agrees with the second term in the velocity expansion
of $D_0$. The series expansion in $\beta$ of 
the scalar function $D_0$ is thus reproduced by that 
of $D_0^h$, while the rest of integration regions does not
contribute at all.
\par
Therefore we have seen, by asymptotic expanding the integral before its
computation, 
that the box amplitude receives
no contributions from the regions of potential, 
soft and ultrasoft loop momentum,
and it is then preserved from Coulomb type singularities, as it was shown
by explicit calculation. 
This fact reveals that, as expected, the box production graph is a 
process dominated by the high scale, as it involves annihilating
photons which carry energies of the order of the mass of the non-relativistic
fermions.
\par
Let us finally note that, although we have reproduced the 
(logarithmic) electron mass
dependence of $D_0$ through the threshold expansion technique, we could, a
priori, need to consider new regions 
to successfully obtain the subleading terms 
${\cal O}(m^2/M^2)\,$, ${\cal O}(m^2/(q^2-4M^2))\,$, etc. This is
what happens, for example, if one considers the 1-loop two-point scalar
function with one heavy mass $M$ and one light mass $m$ at values of 
$q^2\gsim M^2$~: Keeping $m$ finite but smaller than any other
scale present (i.e. 
$m\ll (q^2-M^2)/M\ll\sqrt{q^2-M^2}\ll M$), the integration region 
where $\ell^2\sim m^2$ gives
a non-vanishing contribution proportional to $m^2/(q^2-M^2)$. A new
pattern of integration regimes should then be considered to make each
integral homogeneous also in the $m^2$ scale.     

\section{Conclusions}
\hspace*{0.5cm}
The interest in the study of electron positron annihilation into heavy fermions
has been ushered by the multiple features foreseen both in high--energy
colliders and production at threshold. These include all--important
aspects of the phenomenology like an accurate measurement of the heavy fermion
masses (like $\tau$ or $t$) and, the possibility, of exploring
New Physics beyond the Standard Model.
This goal requires the computation and implementation of complete perturbative
orders within the Standard Model. 
\par
We have evaluated the QED two--photon box diagrams of Fig.~\ref{fig:box} 
contributing to $\sigma (e^+ e^- \rightarrow f \bar{f})$ with massive 
final fermions ($m_e \ll M$), and we have provided a full analytical 
expression for the amplitude. Its contribution at the
production threshold has also been studied and we have found that it is
negligible because of the high velocity suppression. This non--relativistic
analysis complements the one carried out in Ref.~\cite{taus} and shows
that the conclusions reached in that reference are not modified by the
QED box amplitude. 
\par
Finally we have analysed this low velocity behaviour
using the strategy of regions to expand the Feynman integrals near threshold,
confirming that such expansion can also be applied to diagrams involving
heavy and light fermions. This feature allows to identify and 
evaluate, at a fixed order in the heavy fermion velocity, contributions to 
heavy fermion production or annihilation diagrams triggered by light fermions.

\vspace*{1cm} 
\noindent
{\large \bf Acknowledgements}\par
\vspace{0.2cm}
\noindent 
We wish to thank A. Pich for relevant discussions on the subject of this
paper and for a careful reading of the manuscript.
The work of P.~D. Ruiz-Femen\'\i a has been partially supported by a FPU
scholarship of the Spanish {\it Ministerio de Educaci\'on y Cultura}.
J. Portol\'es is supported by a \lq \lq Ram\'on y Cajal" contract with CSIC
funded by MCYT.
This work has been supported in part by TMR, EC Contract No. 
ERB FMRX-CT98-0169, by MCYT (Spain) under grant FPA2001-3031, and
by ERDF funds from the European Commission.

\newpage

\appendix
\newcounter{erasmo}
\renewcommand{\thesection}{\Alph{erasmo}}
\renewcommand{\theequation}{\Alph{erasmo}.\arabic{equation}}
\renewcommand{\thetable}{\Alph{erasmo}}
\setcounter{erasmo}{1}
\setcounter{equation}{0}
\setcounter{table}{0}

\section*{Appendix~A : Integrals in the box amplitude}
\hspace*{0.5cm}
In this Appendix we outline several features of the integration procedure,
followed to evaluate the QED box diagrams, and we collect the explicit
expressions for 
the relevant scalar integrals that appear in our results.
\par
The general structure of the two--photon box amplitude in 
Fig.~\ref{fig:box}(a), ${\cal M}_a$ takes the form 
${\cal M}_a = a_0 D_0 + a^{\mu}D_{\mu} + a^{\mu \nu} D_{\mu \nu}$,
where $a_0, a_{\mu}, a_{\mu \nu}$ contain Dirac algebra $\gamma$'s and spinors,
and $D_0,D_{\mu},D_{\mu \nu}$ are the integrals over the loop momentum 
$\ell$~:
\begin{equation}
D_0;D_{\mu};D_{\mu \nu}  =  \int\frac{d^4\ell}{i(2\pi)^4}\,
\frac{1;\,\ell_{\mu};\,\ell_{\mu}\ell_{\nu}}{(\ell^2-m^2)[(\ell+k-p)^2-M^2]
[(\ell-p)^2-\lambda^2]
[(\ell+p^{\prime})^2-\lambda^2 ]}\,\,,
\label{eq:Ds}
\end{equation}
which depend on three independent four-vectors and where $+i \delta$ 
prescriptions are understood in the propagators . Let us define
our basis as $P=p-k,\, Q=p+p^{\prime}$ and $R=k-k^{\prime}$, with 
scalar products
$$
P^2=t\,\,\,\; \,\,,\,\, \; \,\,\,Q^2=s\,\,\,\; \,\,,\,\,\,\; \,\,R^2=4M^2-s \, ,
$$
$$
P\cdot Q = 0 \,\,\; \,\,\,,\,\; \,\,\,\,P\cdot R=m^2-M^2-t\,\,\; \,\,\,,\,\; \,\,\,\,
Q\cdot R =0\,\,\,.
$$
The integrals in Eq.~(\ref{eq:Ds}) are invariant under the interchange
$\{ p\,;k \} \leftrightarrow \{-p^{\prime}\,;-k^{\prime}\}$. 
Under the same transformation $P\to P$, $Q\to -Q$ and $R\to R$, and thus
the tensor integrals $D_{\mu},D_{\mu \nu}$ do not contain terms linear
in $Q$, justifying our choice of basis. Tensor decomposition of 
$D_{\mu},D_{\mu \nu}$ then reads
\begin{equation}
D_{\mu} =  D_P\, P_{\mu} +D_R \,R_{\mu}
\label{eq:Dvector}
\end{equation}
\begin{equation}
D_{\mu\nu}  =  D_{PP}\,P_{\mu}P_{\nu}+D_{PR}
\Big( P_{\mu}R_{\nu}+ R_{\mu}P_{\nu}\Big)+D_{RR}\,R_{\mu}R_{\nu}+
D_{QQ}\,Q_{\mu}Q_{\nu}+s\,D_{00}\,g_{\mu\nu}\,\,.
\label{eq:Dtensor}
\end{equation}
Further reduction of the coefficient functions appearing in 
Eqs.~(\ref{eq:Dvector},\ref{eq:Dtensor}) has been performed with
the help of $\it FeynCalc$ \cite{FeynCalc}. These coefficients
are thus expressed as a linear combination of a set of
scalar integrals: $D_0,\,C_s,\,C_t,\,C_M,\,B_s,\,B_t$ and $B_M$, with
four ($D_0$), three ($C_a$, $a=s,t,M$) and two ($B_a$, $a=s,t$) propagators
that we collect next.
\par
The relevant scalar integrals have been
evaluated following the method described in \cite{tHooft}, except
for the rather cumbersome 4-point function $D_0$. In the latter case 
we have first calculated its imaginary part in the $s$-channel, 
following the optical theorem, and then the real part has been
reconstructed through the t--fixed unsubtracted dispersion relation that
satisfies $D_0$~:
\begin{equation}
\mbox{Re} D_0(s,t) \, = \, \Frac{1}{\pi} \, \int_{4 \lambda^2}^{\infty}
\! \! \! \!\! \! \! \! \! \! \! {\mbox{\boldmath $-$}} \; \; \;  \, \; 
\, dx \, \Frac{\mbox{Im} D_0(x,t)}{x-s} \; \, ,
\end{equation}
where the Principal Value of the integral is understood.
We have performed its calculation in the 
$\lambda \ll m \ll M$ limit and, therefore, we have neglected 
photon masses when possible. As emphasized in Ref.~\cite{infrared},
the limit $\lambda \to 0$ is not trivial for the occurrence of
terms like $\lambda^2 / (x-4\lambda^2)$, which diverge for finite 
$\lambda$ as $x \rightarrow 4 \lambda^2$ but vanish for $\lambda \to 0$
at fixed $x  \neq 4 \lambda^2$. As a consequence the photon
mass should be kept finite until the final stages.
\par
The scalar integrals that appear in the two--photon box amplitude 
result in Eq.~(\ref{eq:finalM}) through the ${\cal F}_i$ functions
of Eqs.~(\ref{eq:fifunc1}-\ref{eq:fifunc4}) have been evaluated
in the limit where $\lambda \, \ll \,  m \, \ll \, M$ and for the
specific cases $p^2 = p^{\prime 2} = m^2$, $k^2= k^{\prime 2} = M^2$,
$(p+p^{\prime})^2 = (k+k^{\prime})^2 = s$, $(p-k)^2=t$.
They read~:

\begin{eqnarray}
D_0 &=&  \int\frac{d^4\ell}{i(2\pi)^4}\,
\frac{1}{[\ell^2-\lambda^2]\,[(\ell+p)^2-m^2)]\,
[(\ell+p+p^{\prime})^2-\lambda^2]\,
[(\ell+k)^2-M^2 ]}
\nonumber\\[3mm]
&=& \frac{-1}{8\pi^2s\,(M^2-t)}\,\ln  \frac{M^2-t}{mM} \, \ln
\frac{-s-i\delta}{\lambda^2} \, \, ,
\label{eq:D0}\\[6mm]
\overline{D}_0 & = & D_0 \, ( t \rightarrow u ) \, \, ,
\label{eq:D0b}\\[6mm]
C_s &=&  \int\frac{d^4\ell}{i(2\pi)^4}\,
\frac{1}{[\ell^2-\lambda^2]\,[(\ell+p)^2-m^2]\,
[(\ell+p+p^{\prime})^2-\lambda^2]}
\nonumber\\[3mm]
&=& \frac{1}{32\pi^2s} \left[ \,\ln^2\left(\frac{-s-i\delta}{m^2}\right)+
\frac{\pi^2}{3} \,\right] \, \, ,
\label{eq:Cs}\\[6mm]
C_t &=&  \int\frac{d^4\ell}{i(2\pi)^4}\,
\frac{1}{[\ell^2-M^2]\,[(\ell-k)^2-\lambda^2]\,
[(\ell+p-k)^2-m^2]}
\nonumber\\[3mm]
\! &=& \! \! \! 
\frac{-1}{16\pi^2 (M^2-t)}\left[ \,\mbox{Li}_2\left( \frac{t}{M^2}\right)
+ \ln^2\left( \frac{M^2-t}{M m}\right)
+ \ln \left( \frac{M^2-t}{M m} \right) \, \ln \left( \frac{m^2}{\lambda^2}
\right) \, \right] \!  ,
\label{eq:Ct}\\[6mm]
\overline{C}_t & = & C_t \, ( t \rightarrow u ) \, \, ,
\label{eq:Ctb}\\[6mm]
C_M &=&  \int\frac{d^4\ell}{i(2\pi)^4}\,
\frac{1}{[\ell^2-\lambda^2]\,[(\ell+k)^2-M^2]\,
[(\ell+k+k^{\prime})^2-\lambda^2]} 
\nonumber\\[3mm]
&=& \frac{1}{16\pi^2 s\,\beta}\bigg[\,-2\,\mbox{Li}_2(1-\beta)
+2\,\mbox{Li}_2\left( \frac{1-\beta}{1+\beta}\right)
+\frac{1}{2}\ln^2\left(\frac{1-\beta}{1+\beta}\right) 
-2\,\mbox{Li}_2(-\beta)
\nonumber\\[3mm]
&&  \; \; \; \; \; \; \; \; \; \; \; \, \, \; \, \; \; 
-2\ln\beta\,\ln(1+\beta)+i\pi\,
\ln \frac{1-\beta}{1+\beta}\, \bigg] \, \, ,
\label{eq:CM}\\[6mm]
B_s &=&  \int\frac{d^D\ell}{i(2\pi)^D}\,
\frac{1}{[\ell^2-\lambda^2]\,
[(\ell+p+p^{\prime})^2-\lambda^2]} 
\nonumber\\[3mm]
&=& \frac{-1}{16\pi^2}\left(\,\Delta +
\ln\frac {-s-i\delta}{\mu^2}-2\, \right) \, \, ,
\label{eq:Bs}\\[6mm]
B_t &=&  \int\frac{d^D\ell}{i(2\pi)^D}\,
\frac{1}{[\ell^2-M^2]\,
[(\ell+p-k)^2-m^2]} 
\nonumber\\[3mm]
&=& \frac{-1}{16\pi^2}\left(\,\Delta +\ln\frac {-t}{\mu^2}
+\ln\bigg(1-\frac {M^2}{t}\bigg)
-\frac{M^2}{t}\,\ln \frac {M^2-t}{M^2}-2\, \right)
\label{eq:Bt}\\[6mm]
B_M &=&  \int\frac{d^D\ell}{i(2\pi)^D}\,
\frac{1}{[\ell^2-\lambda^2]\,
[(\ell+k)^2-M^2]}
\nonumber\\[3mm]
&=& \frac{-1}{16\pi^2}\left(\,\Delta +\ln\frac {M^2}{\mu^2}
-2\, \right)
\,\,,
\label{eq:BM}
\end{eqnarray} 
where Li$_2(x)$ is the dilogarithm function.
The two--point functions have been
regularized within dimensional regularization in $D$ dimensions and
$\Delta = 2 \mu^{D-4} /(D-4) + \gamma_E - \ln (4\pi)$,
with $\mu$ the renormalization scale. 
From the full expressions above we see that only the integrals
$C_t$, $\overline{C}_t$, $D_0$ and $\overline{D}_0$ are infrared
divergent for vanishing photon mass ($\lambda \to 0$). 
However, as remarked in the main text, the combinations 
$s D_0 - 2 C_t$ (or $s \overline{D}_0 - 2 \overline{C}_t$) are 
infrared finite; accordingly all the infrared divergent contribution
is provided by $D_0$ and $\overline{D}_0$ in Eq.~(\ref{eq:f02IR}) that
carry a $\ln \lambda^2$ factor.

\vspace*{0.6cm}
\newcounter{yago}
\renewcommand{\thesection}{\Alph{yago}}
\renewcommand{\theequation}{\Alph{yago}.\arabic{equation}}
\renewcommand{\thefigure}{\Alph{figure}}
\setcounter{yago}{2}
\setcounter{equation}{0}
\setcounter{figure}{0}

\section*{Appendix~B : Infrared divergence of the QED box diagram}
\hspace*{0.5cm} 
There are several well-known facts on the structure of infrared 
divergences in QED that are relevant for our discussion \cite{IR2001}~:
\begin{itemize}
\item[-] Virtual photon radiative corrections between the external legs 
of a divergenceless root diagram generate an infrared divergent 
contribution that follows a specific pattern in the perturbative
expansion. Such a structure provides a factorization of the resummation
of the divergences at all orders.
\item[-] All the infrared divergence in virtual photon radiative 
corrections commented above, arises from the eikonal approach in the
propagator of the radiating external legs. For spin $1/2$, for example,
with $k$ the outgoing soft photon momentum of $\varepsilon_{\mu}(k)$
 polarization
and $p$ the ingoing external momentum, the modification of the fermion
wave function reads~:
\begin{equation}
u(p) \; \; \; \mapright{photon}{} \; \; \;   
\Frac{1}{p \! \! \!/   \,  - \, 
k \! \! \! /  \, - \, m \, + \, i \, \delta} \; \varepsilon \! \! \! / \; u(p) \, 
= \, \Frac{(2p-k)\cdot \varepsilon \, - \, \frac{1}{2} \,
[k \! \! \! / \, , \varepsilon \! \! \! / \, ]}{k^2 - 2 k\cdot p 
\, + \, i \delta} \, u(p) \; ,
\end{equation}
that, in the eikonal approximation reduces to
\begin{equation}
u(p) \; \; \; \mapright{soft}{photon} \; \; \;  
\Frac{2p \cdot \varepsilon}{k^2 - 2 k \cdot p \, + i \delta} \, u(p)
\; ,
\end{equation}
neglecting, essentially, the spin of the radiating field.
\end{itemize}
Hence to extract the infrared divergent part of the QED box diagram in
Fig.~\ref{fig:box} we need to implement the eikonal approximation into
the amplitude ${\cal M}_a$ in Eq.~(\ref{eq:amplitude}) and the crossed
${\cal M}_b$. This corresponds to evaluate the four diagrams in 
Fig.~\ref{fig:soft}. These are built from the tree--level
diagram for $e^+ e^- \rightarrow f \bar{f}$ through one photon 
annihilation, by attaching a soft photon between an ingoing and
an outgoing external leg in all possible ways. Their 
evaluation gives~:

\begin{figure}[tb]
\begin{center}
\hspace*{-0.5cm}
\includegraphics[angle=0,width=0.65\textwidth]{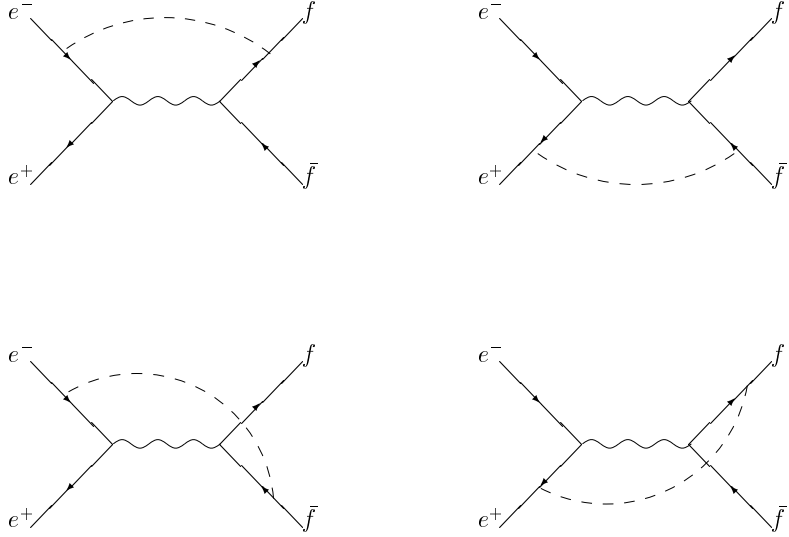}
\caption[]{\label{fig:soft} \it Diagrams contributing to the infrared
divergence of the QED box diagram. The wavy line corresponds to a 
hard photon and the dashed line to a soft photon. As explained in 
the text the infrared divergence factorizes and the spinor structure
is the one of the hard diagram (without radiative corrections).}
\end{center}
\end{figure}

\begin{equation} \label{eq:weinberg}
{\cal M}_{box}^{IR} \, = \, 
\overline{v}_e(p^{\prime}) \, \gamma_{\mu} \,  u_e(p) \, \Frac{e^2 Q_f}{s} \,
\overline{u}_f(k) \, \gamma^{\mu} \, v_f(k^{\prime}) \, 
\left[ \Frac{e^2 Q_f}{4 \pi^2} \,
\ln \left( \Frac{M^2-u}{M^2-t} \right) \, 
\ln \left( \Frac{m^2}{\lambda^2} \right) \, \right] \; ,
\end{equation}
where infrared finite terms have not been written. In fact this result
has been obtained by integrating over the full range of momentum of a massive
photon. Rigorously we should define the infrared contribution by 
imposing an upper limit on its momentum 
$|{\mbox{\boldmath ${q}_{\gamma}$}}| < \Lambda$, and $m^2$ would then be
replaced by $\Lambda^2$ in the logarithm
of ${\cal M}_{box}^{IR}$. In Eq.~(\ref{eq:weinberg}) we have explicitly
stated the factorization between the hard gluon exchange, on the
left,  and the soft
photon exchange inside the square brackets.
\par
Alternatively we can evaluate ${\cal M}_{box}^{IR}$ from our result in 
Eq.~(\ref{eq:finalM}) and we obtain~:
\begin{eqnarray} \label{eq:rawIR}
{\cal M}_{box}^{IR} \,  & = & \,   \Frac{e^4 Q_f^2}{8 \pi^2 \, s} \,  
\left\{  \, \Frac{ (M^2-t) L_1^{- \kappa} \, + \, 2 \, L_2^{+ \kappa} \,
- \, M   \left( L_4^{+ \kappa} - L_4^{- \kappa} \right)}{M^2-t} \, 
\ln \left( \Frac{M^2-t}{M m} \right) \, \right. \nonumber \\[4mm] 
& & \; \; \; \; \; \; \;  \; \; \; \; \; \; 
\left. - \, \Frac{ (M^2-u) L_1^{+ \kappa} \,  - \, 
2 \, L_2^{- \kappa} \, - \, M   \left( L_4^{+ \kappa} - L_4^{- \kappa} 
\right)}{M^2-u} \, \ln \left( \Frac{M^2-u}{M m} \right) \,  
\right\} 
\nonumber \\[4mm]
& & \; \;\; \; \; \; \; \; \; \; \; 
 \times \, \ln \left( \Frac{\lambda^2}{-s-i \delta} \right) 
\, , 
\end{eqnarray}
where the spinor operators $L_i^{\rho \kappa}$ have been defined in 
Eq.~(\ref{eq:spinorbasis}). Then, using the following relations~:
\footnote{Relations between spinor operators like these
can be obtained by explicit evaluation in 
a particular reference frame or transforming the operators into 
traces in the spinor basis, hence working with Lorentz invariant
expressions.}
\begin{eqnarray} \label{eq:tricky}
(M^2-t) \, L_1^{+ \kappa} \, &  = & \, 2 \, L_2^{+ \kappa} \, 
- \, M \, \left( L_4^{+ \kappa} - L_4^{- \kappa} \right) \; \; ,
\nonumber \\[3mm]
(M^2-u) \, L_1^{- \kappa} \, &  = & \, - \,  2 \, L_2^{- \kappa} \, 
- \, M \, \left( L_4^{+ \kappa} - L_4^{- \kappa} \right) \; \; ,
\end{eqnarray}
we finally get~: 
\begin{equation} \label{eq:IRdacord}
{\cal M}_{box}^{IR} \, = \, \Frac{e^2 Q_f}{2 \, s} \, 
\left( L_1^{+ \kappa} \, + \, L_1^{- \kappa} \right) \, 
\left[ \Frac{e^2 Q_f}{4 \pi^2} \, \ln \left( \Frac{M^2-u}{M^2-t} 
\right) \, \ln \left( \Frac{-s-i \delta}{\lambda^2} \right) \right]
\, \, ,
\end{equation}
whose infrared logarithm coincides with our previous result in 
Eq.~(\ref{eq:weinberg}), since $P_{\kappa} u_e(p) = u_e(p)$ in
$L_1^{\pm \kappa}$, being
$\kappa$ the massless electron helicity.
\par
We conclude that the infrared divergence of the QED box
diagram satisfies the expected features \cite{IR2001} and hence its
cancellation should take place when real soft photon radiation 
contributions, at a fixed $\alpha$ perturbative order, are taken into
account.

\end{document}